\documentclass[twocolumn]{aastex631}
\usepackage[]{amsmath,graphicx,natbib,url,multirow,float}

\begin{document}
\newcommand{\kms}{\hbox{km\,s$^{-1}$}}
\newcommand{\Msun}{\mbox{$M_{\sun}$}}
\newcommand{\Lsun}{\mbox{$L_{\sun}$}}
\newcommand{\cm}[1]{\mbox{cm$^{#1}$}}
\newcommand{\cc}{\mbox{cm$^{-3}$}}
\newcommand{\um}{\mbox{$\mu$m}}

\shorttitle{Protosolar D-to-H Abundance and One Part-per-Billion PH$_{3}$ in WISE 0855}
\shortauthors{Rowland et al.}

\title{Protosolar D-to-H abundance and one part per billion PH$_{3}$ in the coldest brown dwarf}
\author[0000-0003-4225-6314]{Melanie J. Rowland}
\affil{University of Texas at Austin, Department of Astronomy, 2515 Speedway C1400, Austin, TX 78712, USA}
\author[0000-0002-4404-0456]{Caroline V. Morley}
\affil{University of Texas at Austin, Department of Astronomy, 2515 Speedway C1400, Austin, TX 78712, USA}
\author[0000-0002-5500-4602]{Brittany E. Miles}
\altaffiliation{51 Pegasi b Fellow}
\affil{Steward Observatory, University of Arizona, Tucson, AZ 85721, USA}
\author[0000-0002-2011-4924]{Genaro Suarez}
\affil{Department of Astrophysics, American Museum of Natural History, New York, NY 10024, USA}
\author[0000-0001-6251-0573]{Jacqueline K. Faherty}
\affil{Department of Astrophysics, American Museum of Natural History, New York, NY 10024, USA}
\author[0000-0001-6098-3924]{Andrew J. Skemer}
\affil{Department of Astronomy and Astrophysics, University of California, Santa Cruz, Santa Cruz, CA 95064, USA}
\author[0000-0002-6721-1844]{Samuel A. Beiler}
\affil{Ritter Astrophysical Research Center, Department of Physics and Astronomy, University of Toledo, 2801 W. Bancroft Street, Toledo, OH 43606, USA}
\author[0000-0002-2338-476X]{Michael R. Line}
\affil{School of Earth  Space Exploration, Arizona State University, Tempe AZ 85287, USA}
\author[0000-0002-9679-4153]{Gordon L. Bjoraker}
\affil{ NASA Goddard Space Flight Center, Greenbelt, MD 20771, USA}
\author[0000-0002-9843-4354]{Jonathan J. Fortney}
\affil{Department of Astronomy and Astrophysics, University of California, Santa Cruz, Santa Cruz, CA 95064, USA}
\author[0000-0003-0489-1528]{Johanna M. Vos}
\affil{School of Physics, Trinity College Dublin, The University of Dublin, Dublin 2, Ireland}

\author[0000-0003-0548-0093]{Sherelyn Alejandro Merchan}
\affil{Department of Astrophysics, American Museum of Natural History, New York, NY 10024, USA}
\affil{Department of Physics, Graduate Center, City University of New York, 365 5th Avenue, New York, NY 10016, USA}
\author[0000-0002-5251-2943]{Mark Marley}
\affil{Lunar and Planetary Laboratory, University of Arizona, Tucson, Arizona, USA}
\author[0000-0003-4600-5627]{Ben Burningham}
\affil{Centre for Astrophysics Research, Department of Physics, Astronomy and Mathematics, University of Hertfordshire, Hatfield AL10 9AB}
\author[0000-0001-9333-4306]{Richard Freedman}
\affil{SETI Institute, Mountain View, CA, USA}
\author[0000-0002-4088-7262]{Ehsan Gharib-Nezhad}
\affil{Space Science and Astrobiology Division, NASA Ames Research Center, Moffett Field, CA, 94035 USA}
\author[0000-0003-1240-6844]{Natasha Batalha}
\affil{Space Science and Astrobiology Division, NASA Ames Research Center, Moffett Field, CA, 94035 USA}
\author[0000-0003-3444-5908]{Roxana Lupu}
\affil{Eureka Scientific, Inc., Oakland, CA 94602}
\author[0000-0001-6627-6067]{Channon Visscher}
\affil{Dordt University, Sioux Center, IA 51250, USA}
\affil{Space Science Institute, Boulder, CO 80301}
\author[0000-0002-6294-5937]{Adam C. Schneider}
\affil{United States Naval Observatory Flagstaff Station: Flagstaff, AZ, US}
\author[0000-0003-2824-3875]{T. R. Geballe}
\affil{Gemini Observatory/ NSF's NOIRLab, Hilo, HI 96720, USA}
\author[0000-0001-5365-4815]{Aarynn Carter}
\affil{Department of Astronomy and Astrophysics, University of California, Santa Cruz, Santa Cruz, CA 95064, USA}
\author[0000-0003-0580-7244]{Katelyn Allers}
\affil{Bucknell University, Lewisburg, PA 17837, USA}
\author[0000-0001-5864-9599]{James Mang}
\affil{University of Texas at Austin, Department of Astronomy, 2515 Speedway C1400, Austin, TX 78712, USA}
\author[0000-0003-3714-5855]{D\'aniel Apai}
\affil{Steward Observatory, University of Arizona, Tucson, AZ 85721, USA}
\affil{Department of Lunar and Planetary Sciences, University of Arizona, Tucson, Arizona, USA}\author[0000-0002-9521-9798]{Mary Anne Limbach}
\affil{Department of Astronomy, University of Michigan, Ann Arbor, MI 48109, USA}
\author[0000-0003-3008-1975]{Mikayla J. Wilson}
\affil{Department of Astronomy and Astrophysics, University of California, Santa Cruz, Santa Cruz, CA 95064, USA}

\correspondingauthor{Melanie Rowland}
\email{mrowland@utexas.edu}

\begin{abstract}
The coldest Y spectral type brown dwarfs are similar in mass and temperature to cool and warm ($\sim$200 -- 400 K) giant exoplanets. We can therefore use their atmospheres as proxies for planetary atmospheres, testing our understanding of physics and chemistry for these complex, cool worlds. At these cold temperatures, their atmospheres are cold enough for water clouds to form, and chemical timescales increase, increasing the likelihood of disequilibrium chemistry compared to warmer classes of planets. JWST observations are revolutionizing the characterization of these worlds with high signal-to-noise, moderate resolution near- and mid-infrared spectra. The spectra have been used to measure the abundances of prominent species like water, methane, and ammonia; species that trace chemical reactions like carbon monoxide; and even isotopologues of carbon monoxide and ammonia. Here, we present atmospheric retrieval results using both published fixed-slit (GTO program 1230) and new averaged time series observations (GO program 2327) of the coldest known Y dwarf, WISE 0855-0714 (using NIRSpec G395M spectra), which has an effective temperature of $\sim$ 264 K. We present a detection of deuterium in an atmosphere outside of the solar system via a relative measurement of deuterated methane (CH$_{3}$D) and standard methane. From this, we infer the D/H ratio of a substellar object outside the solar system for the first time. We also present a well-constrained part-per-billion abundance of phosphine (PH$_{3}$). We discuss our interpretation of these results and the implications for brown dwarf and giant exoplanet formation and evolution.

\end{abstract}

\section{Introduction}\label{introduction} 
Brown dwarfs are the lowest mass product of the stellar initial mass function and typically form like stars via gravitational collapse within molecular clouds \citep{Luhman2007,Bate2019}. This formation mechanism can produce objects with a large range of masses, from hydrogen burning stars with masses $>$ 70 M$_{\rm Jup}$, to deuterium burning objects with masses between $\sim$ 12 and 70 M$_{\rm Jup}$, to objects with masses $<$ 12 M$_{\rm Jup}$ that do not undergo any type of fusion \citep{Morley2024}. Objects with masses $>$ $\sim$ 12 M$_{\rm Jup}$ achieve deuterium fusion in their cores for at least part of their history. In brown dwarfs, deuterium fusion ceases due to either failure to maintain dense and hot enough cores to sustain fusion or exhaustion of the deuterium fuel \citep{Morley2024, spiegel2011}. These deuterium and hydrogen burning limits are metallicity dependent and decrease with increasing metallicity \citep{Morley2024}. In any case, after fusion stops these objects progress through spectral types L, T, and finally Y as they cool \citep{Kirkpatrick2005,Cushing2011}. Because brown dwarfs must have cooled to their current temperatures within a Hubble time, thermo-evolutionary models such as those of \citet{saumonmarley2008} and \citet{phillips2020} suggest that the coldest of the Y dwarfs must be objects with masses $<$ 12 M$_{\rm Jup}$ that never underwent deuterium fusion. 

Observations of deuterium have a long history of informing planet formation, migration, and evolution theories within the solar system starting with the detection and abundance measurement of deuterated methane (CH$_{3}$D) in Jupiter \citep{Beer1972,Beer1973}. Deuterium has the potential to be a mass indicator for larger objects. Nevertheless, there have been no detections of deuterium in any extrasolar atmosphere to date \citep{Morley2019}.  Other isotopes, like those of carbon, oxygen, and nitrogen have been detected in atmospheres of exoplanets and brown dwarfs by several ground- and space-based moderate to high resolution near infrared spectrographs \citep{zhangy2021_isotopebd,zhangy2021_isotopejupiter,line2021Natur,Finnerty2023,gandhi2023,barrado2023, Finnerty2024, smith2024, xuan2024,xuan2024_companions,Hood2024,lew2024, gonzalez2024}. \citet{Molliere2019_isotopes} has proposed that these isotopes can be used as an additional tracer of planet formation for objects in disks, however more work connecting these isotopic species to disk theory needs to be done to determine how these species can trace a object's formation mechanism, formation location, and migration \citep{Bergin2024}. 

\subsection{Deuterium as a mass indicator}
Brown dwarfs are fully convective below their radiative-convective atmospheres \citep{chabrierbaraffe2000, burrows1997}. The presence of deuterated species in the atmosphere means the object did not fuse all of its deuterium, so its presence can be used as a mass indicator.  Evolutionary models by \citet{spiegel2011} and \citet{Morley2024} show that the deuterium burning limit varies with metallicity, helium abundance, and cloud properties, and that objects greater than $\sim$12 M$_{\rm Jup}$ will fuse a fraction ($>$50$\%$) of their deuterium and objects greater than 20 M$_{\rm Jup}$ will fuse all of their deuterium within the first 100 Myr. The presence of deuterium in the atmospheres of all but the youngest objects would indicate that those objects' masses are below the deuterium burning limit. 

\subsection{Deuterium in the solar system}
The ratio of deuterium to hydrogen (D/H) can trace volatile transport within a system in addition to subsequent atmospheric evolution through atmospheric escape. The D/H ratio among comets is enhanced by an order of magnitude over D/H values in the ISM due to the preferred form of water ice being deuterated water (HDO) rather than H$_{2}$O at temperatures below 50 K \citep{cleeves2014}. This enhancement at cold temperatures would also affect the icy mantles of pebbles in the early outer solar system. The D/H ratio of the Earth is similar to that of comets, and \citet{ida2019} showed that Earth's deuterium enhancement could be explained under the pebble accretion theory if some of the pebbles were transported to the inner solar system before the formation of Jupiter opened a gap in the disk and halted pebble migration. \citet{young2023} showed that some of the Earth's water and a fraction of the Earth's deuterated water could also be formed via room temperature reactions between FeO in an early magma ocean and a primordial hydrogen-dominated atmosphere.  This would suggest that Earth's deuterium abundance is due to multiple processes and only partially attributable to the transport of volatiles within a system. 

The D/H ratios in the Martian and Venutian atmospheres are enhanced by several orders of magnitude over the Earth's value because of the preferential depreciation of lighter hydrogen atoms over heavier deuterium atoms in atmospheric loss processes \citep{drake2005}. The D/H ratio of the ice giants is also enhanced because they accreted deuterium-enriched ices during formation and large fractions of their envelopes are comprised of these ices \citep{Feuchtgruber2013}. While the gas giants Saturn and Jupiter also accreted these ices, their envelopes are dominated by helium and molecular hydrogen, so no enrichment is detected above the protosolar ratio (1-2 $\times$ 10$^{-5}$) \citep{Lellouch2001}. 

\subsection{Phosphine in substellar objects}
Phosphine (PH$_{3}$) is the observed reservoir of phosphorous in gas giant planets in the solar system and is the expected reservoir of phosphorous in giant extrasolar atmospheres colder than $\sim$1000 K \citep{Ridgway1976, Larson1977, Larson1980}.  PH$_{3}$ has a significant absorption feature at $\sim$4.2 $\mu$m, but has been difficult to detect at expected quantities in all cold atmospheres outside of the solar system \citep{visscher2006, Morley2018, Hood2024, faherty2024, kothari2024, beiler2024b}. Vertical mixing in cold atmospheres is expected to bring PH$_{3}$ from the warm interior into the photosphere. This effect has been seen on Jupiter (T$_{\rm eff}$ = 125 K) which has a PH$_{3}$ abundance of 1-2 ppm (10$^{-5.7}$--10$^{-6}$) \citep{fletcher2009}. However, outside of a tentative detection from \citet{burgasser2024},  PH$_{3}$ has remained elusive in moderate and high-resolution spectra of T dwarfs and low resolution spectra of Y dwarfs down to T$_{\rm eff}$ = 264 K. In particular, vertical mixing rates inferred from measured CO abundances imply large amounts of PH$_{3}$ should be detected in cold atmospheres \citep{miles2020}. However, \citet{Morley2018} placed an upper limit of $<$ 10$^{-6.30}$ on the PH$_{3}$ abundance in the coldest brown dwarf, which is orders of magnitude less than the expected value. 

\subsection{WISE 0855}
WISE J085510.83-071442.5 (hereafter WISE 0855) is the coldest known brown dwarf with T$_{\rm eff}$ = 264 K and an estimated mass of 3 M$_{\rm Jup}$ $<$ M $<$ 10 M$_{\rm Jup}$ \citep{Luhman2014,Esplin2016,Leggett2017,Leggett2021, Luhman2024}. It is the fourth closest stellar or brown dwarf system to the Sun with a distance of just 2.28 pc \citep{Kirkpatrick2021}. Its Earth-like temperature and inferred planetary mass has made it a useful analogue of temperate gas giant planets that are lacking in the solar system, since due to their lower masses Jupiter and Saturn have both cooled to $<$ 150 K. \citet{skemer2016} suggested that deuterium could be detected in the form of CH$_{3}$D and \citet{Morley2019} calculated that a 10$\sigma$ detection of CH$_{3}$D is possible with less than 2.5 h of observations by the James Webb Space Telescope's (\textit{JWST}) moderate resolution (R $\sim$ 2700) G395H mode. 

Here we present detections of deuterated methane (CH$_{3}$D) and phosphine (PH$_{3}$) in WISE 0855 using two independent \textit{JWST} NIRSpec/G395M (R $\sim$ 1000) observations. These are the first simultaneous detection of deuterium and the first abundance measurement of PH$_{3}$ in an extrasolar atmosphere. We quantify the effective mixing timescale for PH$_{3}$ and calculate the D/H ratio for WISE 0855.

\section{Methods}\label{Methods}

\subsection{Observations and Data Reduction}\label{Data}
Two independent WISE 0855 data sets reduced by different teams were used for the retrieval analysis. The first data set was taken as part of \textit{JWST} Guaranteed Time Observation (GTO) program 1230 (PI: Alves de Oliveira) and published by \citet{Luhman2024}. The details of that program and the reduction are provided in \citet{Luhman2024}. Program 1230 observed WISE 0855 in fixed slit mode of JWST/NIRSpec with the G395M/F290LP grating/filter setting and a three-point dither pattern. The total exposure time was 15,200 s. The second data set is from General Observer (GO) program 2327 (PI: Skemer, Co-PIs: Morley and Miles). Program 2327 used the Bright Object Time Series (BOTS) mode of JWST/NIRSpec with the G395M/F290LP grating/filter combination and no dithering over 11 h. We discuss the data reduction of program 2327 and differences with program 1230 in this section.

The weighted average spectrum covers 2.87–5.10 $\mu$m at a resolution of $\sim$1000. The observations started on December 02, 2023 at 01:03:18.92 UTC and ended December 02, 2023 at 12:09:33.18 UTC. The 11-hour total exposure time was composed of forty-four, 15-minute integrations. The observations were reduced using version 1.14.0 of the standard JWST Pipeline \citep{bushouse_2024_12692459} with CRDS version `11.17.20' and CRDS context `\texttt{jwst\_1215.pmap}'. Stage 1 was run with the default parameters to correct detector level artifacts and convert raw detector images into slope images. Stage 2 of the pipeline corrects residual detector artifacts at the integration level and converts slope images into flux calibrated 2-dimensional spectral images. All default Stage 2 steps are run with one additional step. The \texttt{nsclean} step was turned on to remove correlated read noise. The 2-dimensional individual spectral images are then used for spectral extraction.

The 44 spectral images were split into six separate segment files. The first 5 segments each hold 8 spectral images, the last segment has 4 spectral images. The average 2-dimensional spectral image is calculated for each segment and used to estimate the shape of the spectral trace. At every column (y-dimension) a 1-dimensional Gaussian is fit to sub-pixel precision to estimate the center (in x-dimension) of the trace. The x- and y- values are used to fit a second order Chebyshev series to the spectral trace\footnote{Example - \url{https://github.com/exonik/JWebbinar2023-TSO/blob/main/Part2-Spec2.ipynb}}. We defined an 8-pixel wide extraction aperture centered at the best fit at each column. The excess background at each column is estimated by taking a median of the 3-pixels outside of the extraction aperture from both sides. The median excess background is then subtracted off from the extracted spectrum. 

\begin{figure*}
    \centering
    \includegraphics[width=1.0\linewidth]{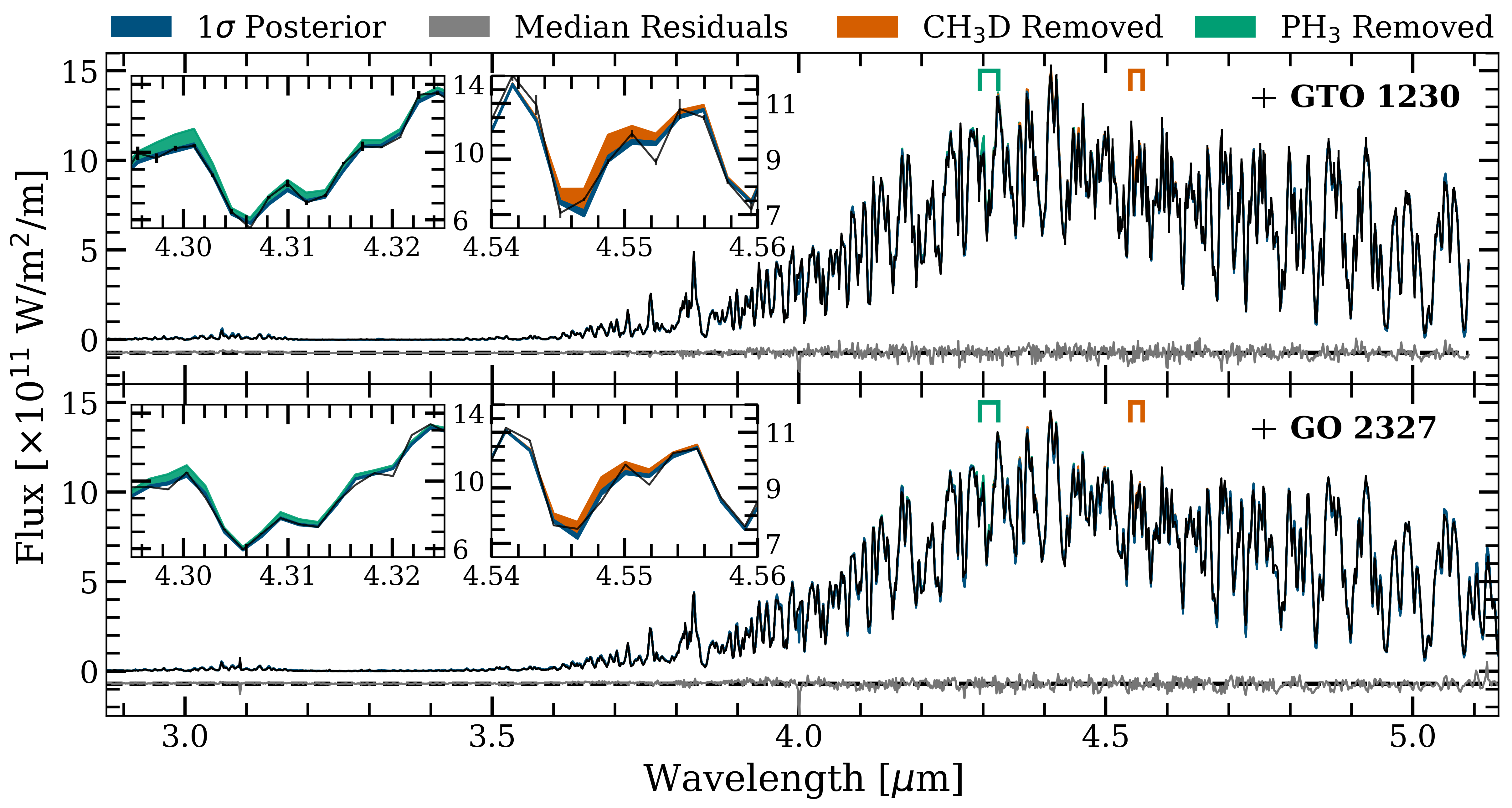}
    \caption{The JWST/NIRSpec G395M spectrum with error bars of WISE 0855 for GTO program 1230 (PI: Alves de Oliveira) [top] and GO program 2327 (PI: Skemer) [bottom], 1$\sigma$ retrieved spectrum, and the residuals from the median retrieved spectrum are shown in black, blue, and grey. The difference between the best fit spectrum and the best fit spectrum with PH$_{3}$ or CH$_{3}$D opacity removed are plotted in shaded green and orange, respectively. The regions where each opacity has the largest effect on the spectrum ($\approx$ 10$\%$) are highlighted in sub-panels, showing that both PH$_{3}$ or CH$_{3}$D are detected in both datasets. The GTO dataset has a mean S/N of 107 and the GO data set has a mean S/N of 759.}
   \label{fig:spectra}
\end{figure*}

Each spectral image has an associated error image and wavelength map. The same extraction radius is used to estimate the error using standard error propagation and wavelength solution of each pixel. After all 44 spectra were extracted, they were visually compared to mask hot pixels and other outliers. The weighted average and propagated errors of all the masked spectra were calculated to produce a high signal-to-noise (S/N) spectrum of WISE 0855 for retrieval analysis shown in Figure~\ref{fig:spectra}. The {\it JWST} data used in this analysis can be found in MAST: \dataset[10.17909/rjtp-zn54]{http://dx.doi.org/10.17909/rjtp-zn54}.

Understanding the differences between the spectrum presented in \cite{Luhman2024} and this work is challenging due to WISE 0855's inherent variability and the $\sim$9 month time difference between observations. Finding the differences between the pipeline used for the \cite{Luhman2024} spectrum and the standard JWST pipeline would be ideally addressed with non-variable, standard sources, but is beyond the scope of this work. For the purposes of the retrieval analysis, the reductions from \cite{Luhman2024} and this work are treated as two different epochs. Additional discussion of the reductions are included in Appendix~\ref{AppendixA}.

\subsection{Retrieval}\label{RetreivalMethod}
We performed a suite of retrievals on the spectra using the GPU-enabled CHIMERA retrieval framework \citep{line2015}. All models used the radiative transfer code described in \citet{Hood2023}, which adapted the GPU-enabled radiative transfer from \citet{zalesky2022} to solve the two-stream multiple scattering problem using the methods from \citet{toon1989}. We utilized the Anaconda Numba \texttt{guvectorize} framework on NVIDIA A100 GPUs.  The GPU memory (40 GB) limited the number of simultaneous CPU threads to 4. Like the previously mentioned studies, parameter estimation was conducted using the \texttt{emcee} package \citep{foremanmackey2013}. All retrievals used a minimum of 8 walkers per parameter and were run to 60,000 iterations and took approximately 40 h. Initial \texttt{emcee} walker positions in parameter space were constructed using a Gaussian ball centered on by-eye fits informed by Sonora Elf Owl models \citep{mukherjee2024}. 

We modified the free temperature-pressure profile described in \citet{line2015} to directly retrieve the temperature at 18 points equally spaced in log$_{10}$ pressure between $-$4.3 and 2.5 to ensure the capture of any potential temperature inversion above 1 millibar as seen in {\citet{faherty2024}}. 

We included the following gas opacities: H$_{2}$O \citep{polyansky2018}, CH$_{4}$ \citep{Hargreaves2020ApJS}, CH$_{3}$D \citep{Hargreaves2020ApJS}, $^{12}$CO \citep{rothman2010,li2015}, $^{13}$CO \citep{rothman2010,li2015}, CO$_{2}$ \citep{Huang2014}, NH$_{3}$ \citep{Yurchenko2011}, H$_{2}$S \citep{Tennyson2012, azzam2015}, and PH$_{3}$ \citep{Sousa-Silva2015,Tennyson2012} and H$_{2}$-H$_{2}$ and H$_{2}$-He collision induced opacities. We did not include HDO because its expected opacity is several orders of magnitude below the dominant opacity sources at all observed wavelengths.  We assume uniform-with-altitude mixing ratios. We include non-grey H$_{2}$O clouds using the \texttt{EDDYSED} model with a log-normal particle size distribution and 3 retrieved values (cloud base pressure, sedimentation efficiency f$_{\rm sed}$, and cloud volume mixing ratio) \citep{ackermanmarley2001}. Additional retrievals with a non-uniform-with-altitude mixing ratio for H$_{2}$O were performed using the method described in \citet{Rowland2023}. CH$_{3}$D cross sections were scaled to terrestrial abundances (6.227 $\times$ 10$^{-4}$ or 1:1606 relative to CH$_{4}$). We retrieve the log of the CH$_{3}$D abundance relative to the terrestrial CH$_{3}$D/CH$_{4}$ ratio. 

Models were run at a variable resolution ranging from R $\sim$ 18,000 at shorter wavelengths to R $\sim$ 35,000 at longer wavelengths to ensure a constant 25 model points per instrumental resolution across the entire wavelength range (2.9 - 5.55 $\mu$m). Model spectra were rotationally broadened according to a \textit{v} sin \textit{i} parameter, Doppler-shifted according to a radial velocity parameter, convolved to the instrumental resolution of G395M ($\sim$2.2 pixels/resolution element), and binned.

Additional retrievals with different chemistry profile, temperature profile, cloud, and rotational broadening treatments were performed. All parameters and their prior ranges are provided in Table \ref{tab:params} and a list of all retrievals performed is provided in Table \ref{tab:retrievals}. Some parameters (e.g., surface gravity, radius, and abundance profiles) were retrieved directly while other bulk properties like the P(T) profile, effective temperature, metallicity, and C/O were derived based on retrieved parameters. These derivations are described below. 

To calculate the effective temperature, we equate Boltzmann's law to the bolometric flux between 0.77 and 30 $\mu$m, similar to the method used in \citet{line2017}. 

Metallicity is computed as, 
\begin{equation}
\left[ \rm M/\rm H \right] = \rm log_{10}\left( \frac{(M/H)_{retrieved}}{(M/H)_{solar}}\right)
\label{eq:metallicity}
\end{equation}
where the retrieved metallicity is taken to be the summation of the elemental species included in the retrieval model. The C/O ratio is computed as, 
\begin{equation}
\frac{\rm C}{\rm O} = \frac{\sum \rm C}{\sum \rm O} \approx \frac{\rm CH_{4} + \rm CH_{3}D + \rm ^{12}CO + \rm ^{13}CO + \rm CO_{2}}{1.22 \times(\rm H_{2}O + \rm ^{12}CO + \rm ^{13}CO + \rm 2CO_{2})}
\label{eq:coratio}
\end{equation}
\citet{calamari2024} found a median atmospheric oxygen sink of 17.8$\%$$^{+1.7\%}_{-2.3\%}$ for brown dwarfs in the solar neighborhood due to oxygen sequestration into silicate clouds deep below the photosphere. This corresponds to an oxygen correction of  1.22$^{+0.02}_{-0.04}$, as shown in Equation \ref{eq:coratio}. This correction does not account for any oxygen that may be sequestered in water clouds. 

CH$_{3}$D was retrieved as the log$_{10}$ ratio of CH$_{3}$D/CH$_{4}$ relative to the terrestrial ratio (1:1606), with a broad prior of -3 to 3 (corresponding to CH$_{3}$D abundance 1/1000 to 1000 $\times$ terrestrial). To calculate the abundance of CH$_{3}$D, we used the following equation:

\begin{equation}
f_{\rm CH_{3}D} = f_{\rm CH_{4}} \times C_{E} \times f_{\rm CH_{3}D/\rm CH_{4}}
\label{eq:ch3dabundance}
\end{equation}
where $f_{CH_{3}D}$ is volume mixing ratio (VMR) of CH$_{3}$D, $f_{\rm CH_{4}}$ is the VMR of CH$_{4}$, $C_{E}$ is 6.227 $\times$ 10$^{-4}$ (the CH$_{3}$D/CH$_{4}$ of Earth), and $f_{\rm CH_{3}D/CH_{4}}$ is the unlogged retrieved ratio.
The D/H ratio as measured in CH$_{3}$D is computed as 

\begin{equation}
(\rm D/\rm H)_{\rm CH_{4}} = \frac{\sum \rm D}{\sum \rm H} \approx \frac{\rm CH_{3}D}{\rm 4CH_{4} + \rm 3CH_{3}D}
\label{eq:dhratio}
\end{equation}

The D/H ratio of H$_{2}$ in the envelope, (D/H)$_{\rm H_{2}}$, is taken to be the bulk D/H ratio of the object. This assumption is valid if the components were well mixed at least once in the object's history, as would be the case if WISE 0855 is a low mass product of the initial mass function. However, the D/H ratio is not inferred from a HD/H$_{2}$ measurement, but by the CH$_{3}$D/CH$_{4}$ measurement, (D/H)$_{\rm CH_{4}}$. The chemical reaction that produces CH$_{3}$D in an atmosphere is:

\begin{equation}
\rm CH_{4} + \rm HD \leftrightarrow \rm CH_{3}D + \rm H_{2}
\label{eq:ch3dreaction}
\end{equation}

The fractionation factor that describes the ratio of (D/H)$_{\rm CH_{4}}$ to (D/H)$_{\rm H_{2}}$ is described in \citet{lecluse1996_fractionation}. It decreases with increasing CH$_{3}$D quench temperature until reaching a value of 1.0 at quench temperatures $>$ 1200 K. Jupiter has a CH$_{3}$D quench temperature of 790 K and a measured fractionation factor of 1.25, with lower quench temperatures and higher fractionation factors for the colder planets \citep{lecluse1996_fractionation,FegleyPrinn_1988_fractionation}. We do not have a robust constraint on the CH$_{3}$D quench temperature in WISE 0855 so we estimate a fractionation factor between 1.0 and 1.1 based on extrapolations from the solar system fractionation factors.

The inclusion of each trace gas species adds one parameter to the model. To assess whether the additional parameter is warranted (i.e. improves the model fit to the data), we computed the Bayesian information criterion (BIC) for each retrieval. The BIC was computed as 
\begin{equation}
BIC = -2 \ln{(L)} + \ln{(N)}K 
\label{eq:bic}
\end{equation}
where $\ln{(L)}$ is the log-likelihood of the best-fit model, N is the number of data points, and K is the number of parameters. The model with the lower BIC indicates the better model. We select between two models using the following intervals used in \citet{kassraftery1995} with evidence against the higher BIC as 0 $<$ $\Delta$BIC $<$ 2: no preference worth mentioning; 2 $<$ $\Delta$BIC $<$ 6: positive; 6 $<$ $\Delta$BIC $<$ 10: strong; and 10 $<$ $\Delta$BIC: very strong.

\section{Results}\label{Results}
Spectra retrieved for both the GTO program 1230 (PI: Alves de Oliveira, hereafter ``GTO data") data set published by \citet{Luhman2024} and the averaged time series observations from GO program 2327 (PI: Skemer, hereafter ``GO data") are presented in Figure \ref{fig:spectra}.

\begin{figure*}
    \centering
    \includegraphics[width=0.675\linewidth]{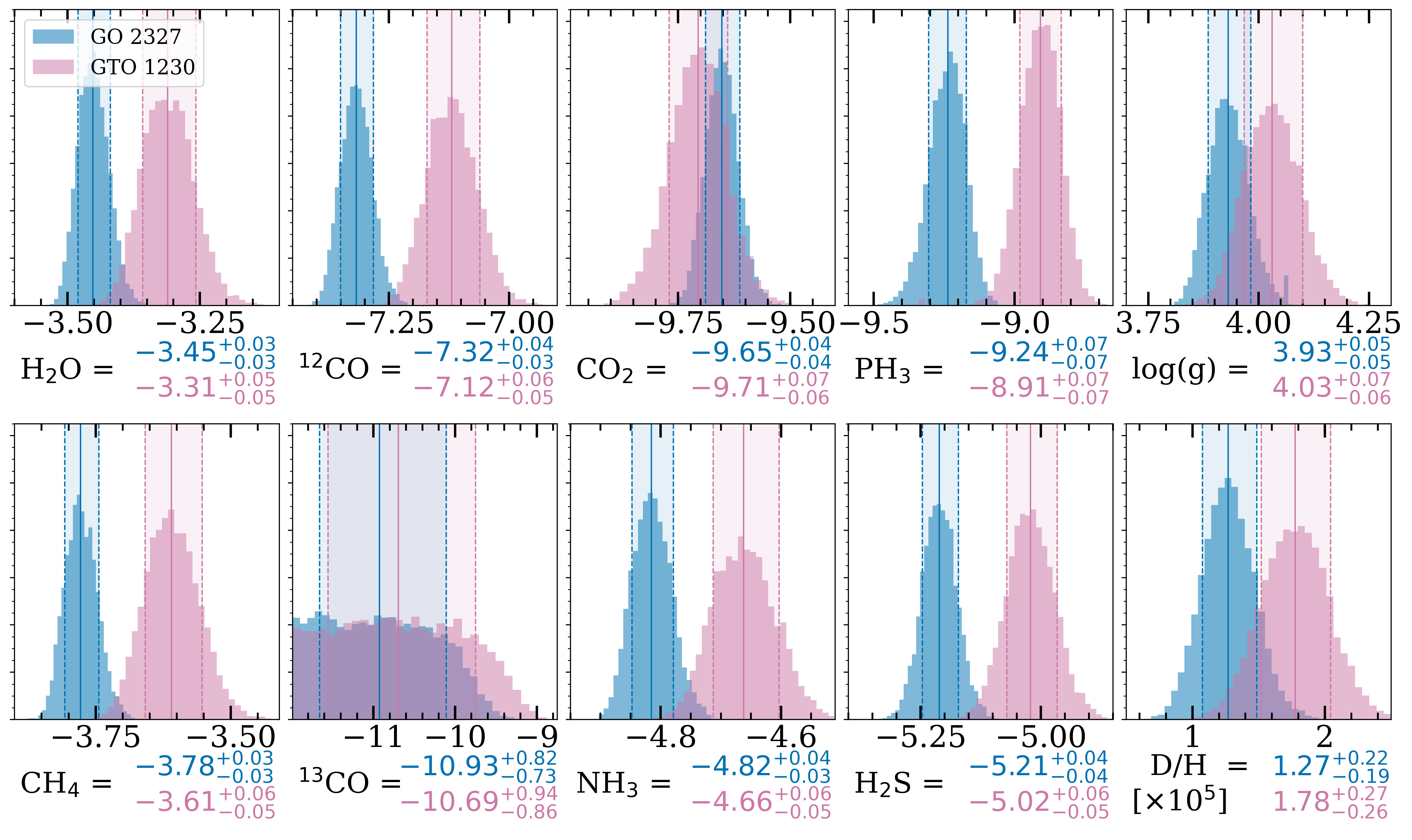}
    \includegraphics[width=0.315\linewidth]{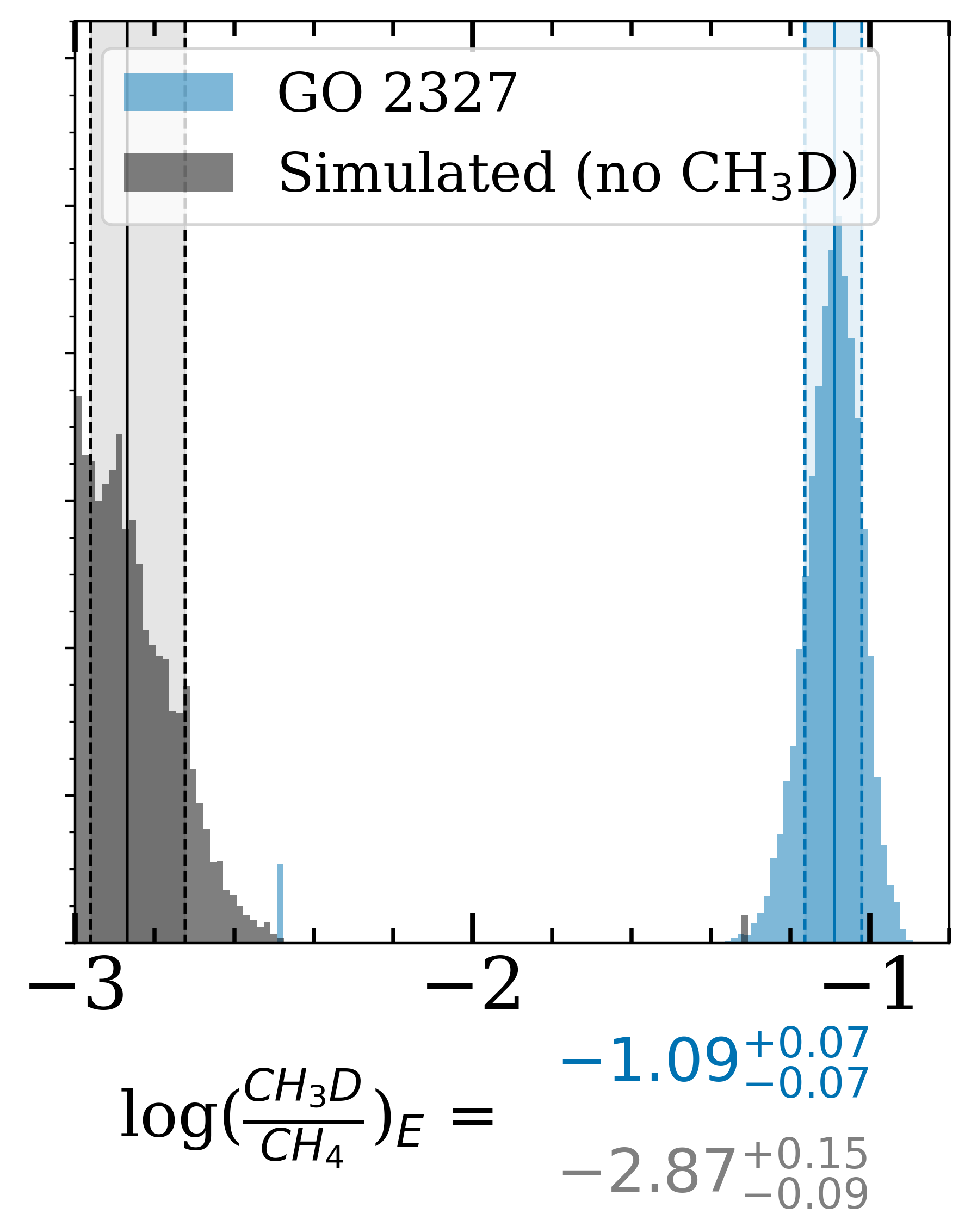}
    \caption{Left panel: The retrieved posteriors for the gas-phase abundances and log(g) for the GTO spectrum [pink] and GO spectrum [blue]. All gases are constrained with the exception of $^{13}$CO. Right Panel: The retrieved log($\frac{\rm CH_{3}D}{\rm CH_{4}}$)$_{\rm E}$ posterior from the GO spectrum [blue] and a simulated spectrum based on the best fit model with CH$_{3}$D opacity set to 0. The bounded constraint from the actual data and the upper limit from the synthetic data indicates that CH$_{3}$D information exists in NIRSpec/G395M data.} 
   \label{fig:ccf}
\end{figure*}

\begin{figure*}
    \centering
    \includegraphics[width=0.42\linewidth]{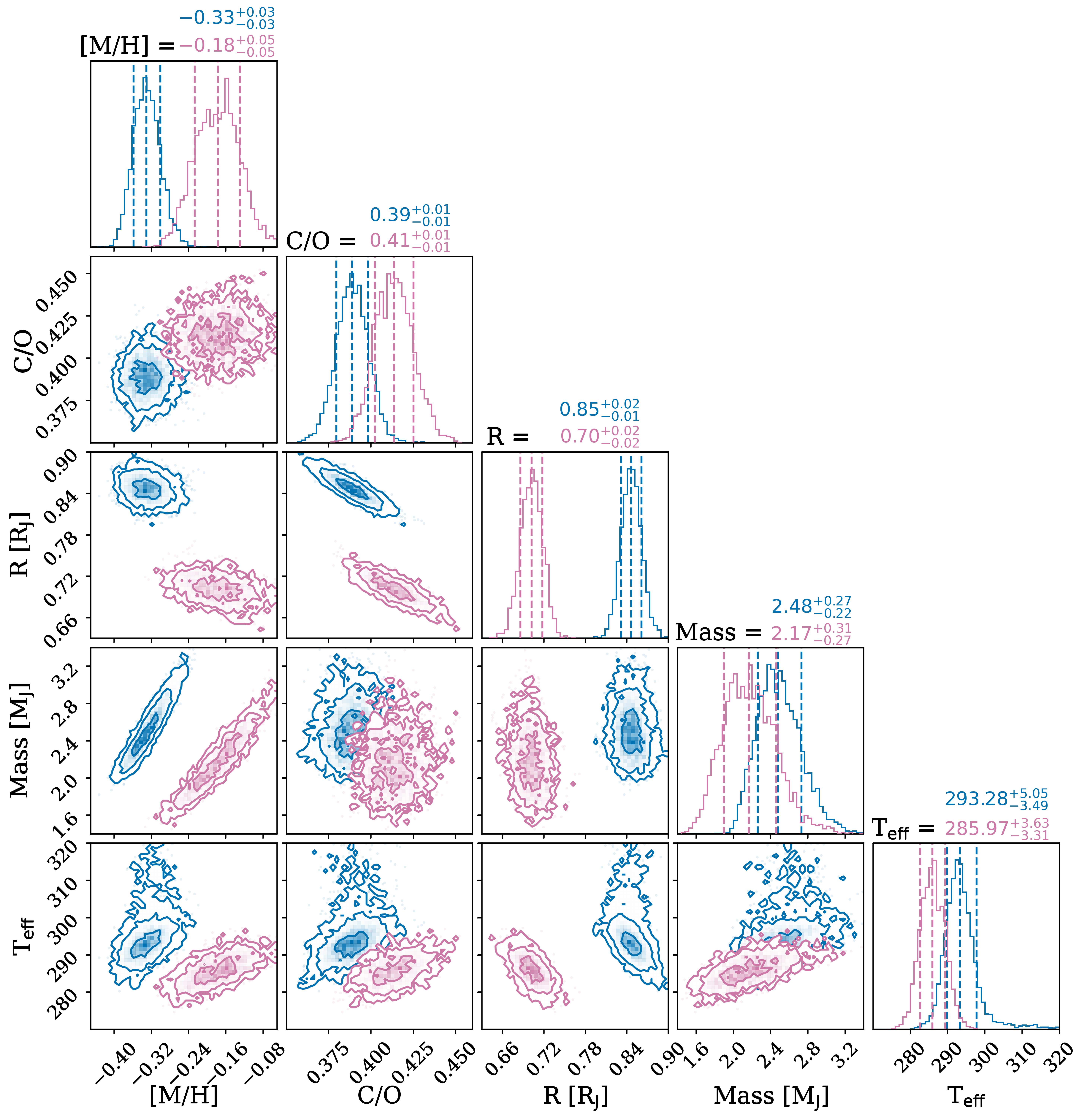}
    \includegraphics[width=0.57\linewidth]{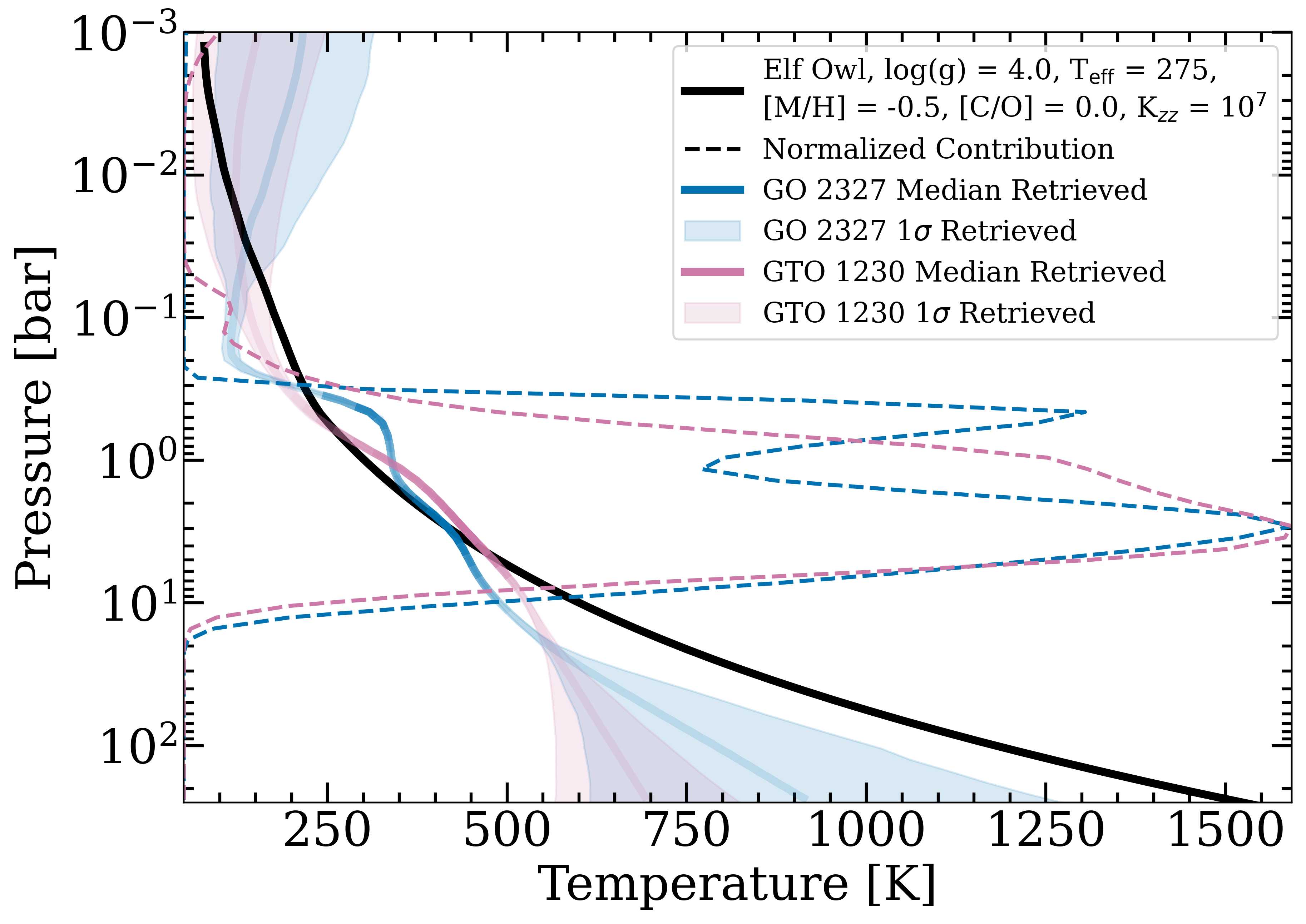}
    \caption{The corner plot of derived parameters and the and P(T) profile of the fiducial retrieval on the GTO spectrum [pink] and the time-averaged GO spectrum [blue]. The dashed lines in the P(T) plot show the normalized flux averaged contribution for each layer in the atmosphere and the opaqueness of the P(T) profile corresponds to this value, with a minimum of 20$\%$ for visibility.}
   \label{fig:corner}
\end{figure*}

Retrieved gas abundances and log(g) are shown in Figure \ref{fig:ccf} and the derived parameters and the P(T) profiles are shown in Figure \ref{fig:corner}. While the retrieved P(T) profiles differ from the self consistent Elf Owl model, they generally agree with each other and the self consistent model in the photosphere. Both retrievals show these deviations despite a smoothing hyperparameter which may be due to inhomogeneities like patchy cloud cover or other 3-dimensional effects.  

We calculated the bolometric luminosity (L$_{\rm bol}$) using the PRISM spectrum published in \citet{Luhman2024} and the method described in \citet{Beiler2024a} and determine a log L/L$_{\sun}$ = -7.297 $\pm$ 0.042, which agrees with the log L/L$_{\sun}$ = -7.305 $\pm$ 0.020 calculated in \citet{Luhman2024}. We calculated a T$_{\rm eff}$ = 264 $\pm$ 8 K using an assumed age of 1 to 10 Gyr to estimate the radius from evolutionary models \citep{marley2021ApJ...920...85M}, which is slightly colder than the T$_{\rm eff}$ = 285 K determined by model fitting in \citet{Luhman2024}.

In the GO data set (GTO data set) we retrieve subsolar metallicity [M/H] = -0.33 $\pm$ 0.03 (-0.18$\pm$ 0.05), C/O ratio of 0.39 $\pm$ 0.01 (0.41 $\pm$ 0.01) compared to a solar C/O of 0.48, and log(g) = 3.93 $\pm$ 0.05 (4.03 $\pm$ 0.07). We find that G395M spectra do not have sufficient resolution to constrain \textit{v} sin \textit{i}. We derive a slightly higher T$_{\rm eff}$ = 293$^{+5}_{-3}$ K (286$^{+4}_{-3}$ ) compared to the T$_{\rm eff}$ expected from L$_{bol}$ (264$\pm8$). We derive a mass of 2.48$^{+0.27}_{-0.22}$ M$_{\rm Jup}$ (2.17$^{+0.31}_{-0.27}$ M$_{\rm Jup}$). Both the high T$_{\rm eff}$ and low mass are driven by the low retrieved R = 0.85$^{+0.02}_{-0.01}$ $R_{\rm Jup}$ (0.70 $\pm$ 0.02), a common problem with brown dwarf retrievals \citep{Zalesky2019, Gonzales2020, kitzmann2020heliosr2, Burningham2021}. \citet{Luhman2024} estimated a radius of 0.9 $R_{\rm Jup}$ and evolutionary models predict a radius between 0.97 and 1.1 $R_{\rm Jup}$ \citep{marley2021ApJ...920...85M}. Assuming an R = 1.0 $R_{\rm Jup}$, we recalculate a mass of 3.44 M$_{\rm Jup}$ (4.33 M$_{\rm Jup}$) based on the retrieved scale factor and log(g).

While many of the bulk properties retrieved in each data set qualitatively agree, we note that some parameters, like some gas abundances and the radius, disagree by more than 1$\sigma$. We note that abundance ratios tend to be more robust than individual abundances, with our retrieved C/O and CH$_{3}$D/CH$_{4}$ ratios agreeing within 1$\sigma$. The difference in radius may be due to differences in the data reduction pipelines used for each data set. WISE 0855 is a variable object and observations were taken 9 months apart. If this variability is driven by clouds, clouds may impact the gas-phase abundances of prominent opacity sources like H$_{2}$O. Finally, the GTO observations spanned 1 h and the GO observations spanned 11 h and covered all or most of a rotation period. If spacial or temporal inhomogeneities exist, they may have impacted each data set differently. The impact of these factors on retrieved posteriors will be explored in a future work.

\subsection{\texorpdfstring{CH$_{3}$D }D detection}
Using the high S/N time series observations, we retrieve a $\log_{10}$ CH$_{3}$D/CH$_{4}$ of -1.09 $^{+0.06}_{-0.07}$ relative to terrestrial ratios, which corresponds to a (D/H)$_{\rm CH_{4}}$ of 1.27 $^{+0.22}_{-0.19}$ $\times$ 10$^{-5}$, and a bulk D/H ratio of 1.15 $^{+0.34}_{-0.17}$ $\times$ 10$^{-5}$, assuming a fractionation factor between 1.0 - 1.1. Using the GTO observations, we retrieve a slightly larger (D/H)$_{\rm CH_{4}}$ of 1.78 $^{+0.27}_{-0.26}$ $\times$ 10$^{-5}$ and a bulk D/H of 1.62 $^{+0.43}_{-0.24}$ $\times$ 10$^{-5}$. These values agree within 1$\sigma$ of the GTO results and are roughly consistent with a protosolar ratio of 1-2 $\times$ 10$^{-5}$. These retrieved values are robust to all model parameterizations tested and are listed in Appendix \ref{AppendixB}. The retrieved D/H values were invariant to walkers initialized with Gaussian balls centered on a terrestrial abundance (a high CH$_{3}$D/CH$_{4}$ abundance) and one one-thousandth a terrestrial abundance (a very low CH$_{3}$D/CH$_{4}$ abundance). The retrieved D/H value did not change within the 1$\sigma$ errors for retrievals with H$_{2}$O Mie scattering clouds or no clouds, with uniform-with-pressure chemistry abundance profiles for all gases (including H$_{2}$O), and with a non-uniform profile for H$_{2}$O,  with a smoothed or unsmoothed P(T) profile, or with the \texttt{FastRotBroad} and the \texttt{Rot Broad Int} rotational broadening functions. 

We performed retrievals without CH$_{3}$D to test the strength of our detection. The models with CH$_{3}$D opacity were strongly preferred with a $\Delta$BIC value of -44.9 in the GO data set and a $\Delta$BIC value of -55.4 in the GTO data set. 

To bolster confidence in the CH$_{3}$D detection, we performed an injection and retrieval test similar to the method in \citet{line2021Natur}. We created a synthetic data set using our best fit model from the GO data set retrieval but with CH$_{3}$D opacity removed. We added synthetic noise by sampling each data point from a normal distribution characterized by the error bar at each wavelength. We then performed a retrieval including CH$_{3}$D as a parameter and only retrieve an upper limit. The results are shown in Figure \ref{fig:ccf}. The bounded constraint from the actual data and the upper limit from the synthetic data indicates that CH$_{3}$D information exists in NIRSpec/G395M data and is sufficient to constrain the CH$_{3}$D abundance. 

\subsection{\texorpdfstring{PH$_{3}$ d}  detection}
We also detect PH$_{3}$ in both data sets. In the GO data set we detect a $\log_{10}$(VMR) abundance of -9.24 $\pm$ 0.07 and in the GTO data set we detect a $\log_{10}$(VMR) abundance of -8.91 $\pm$ 0.07. Similarly to the CH$_{3}$D tests described in the previous section, we performed retrievals with and without PH$_{3}$. The models with PH$_{3}$ were strongly preferred over those without with a $\Delta$BIC value of -27.9 in the GO data set and a $\Delta$BIC value of -61.9 in the GTO data set. We performed a cross-correlation analysis following the methods of \citet{zhangy2021_isotopebd}, in which the residuals of the best fit model with PH$_{3}$ opacity removed and a PH$_{3}$ model are cross-correlated.  This resulted in a CCF S/N of 8.5. The results of this analysis are included in Appendix \ref{AppendixB}.

\section{Discussion}\label{Discussion}
\begin{figure}
    \centering
    \includegraphics[width=1.0\linewidth]{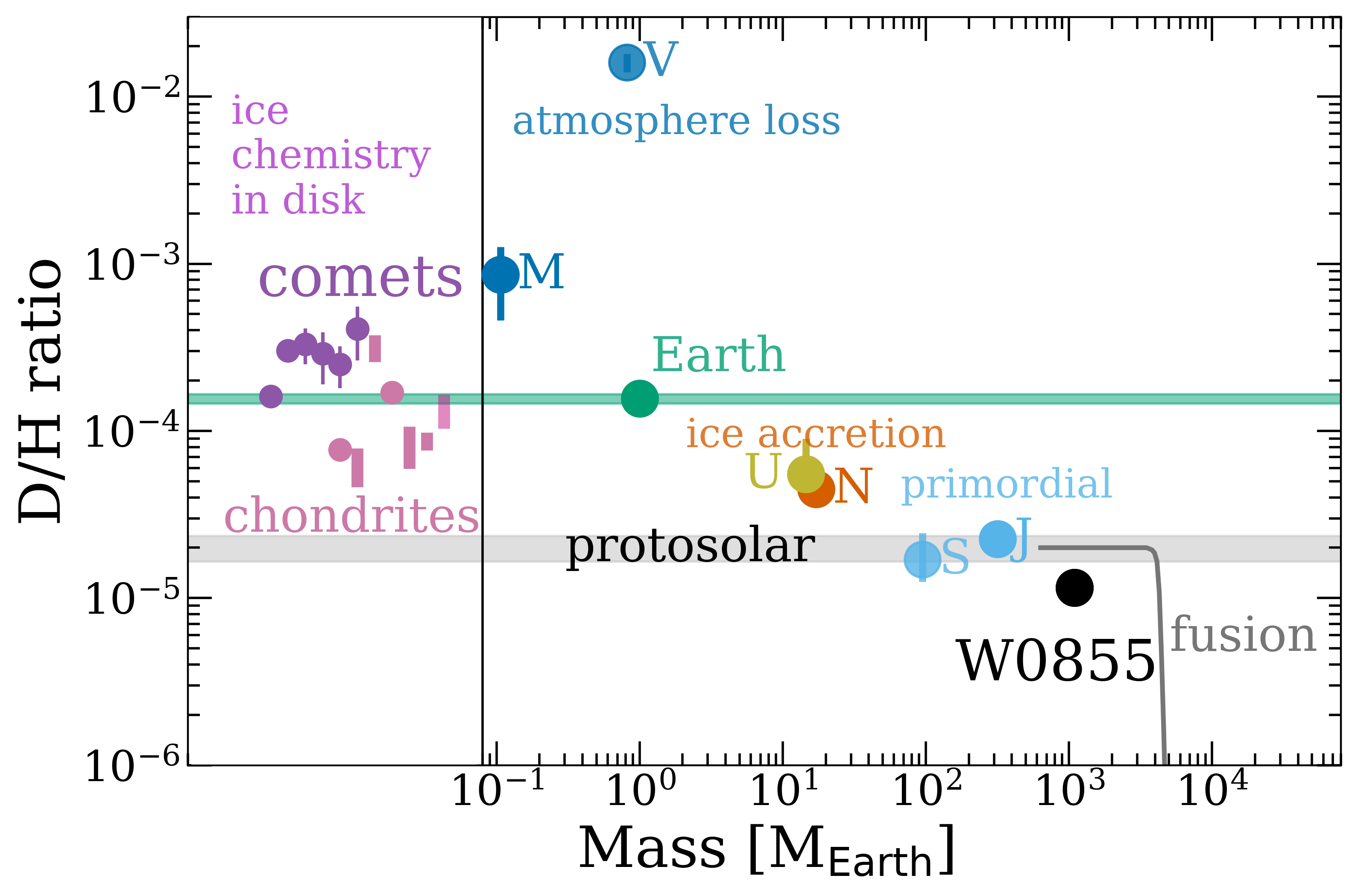}
    \caption{D/H ratio as a function of object mass as measured in the ISM (grey), terrestrial planets, ice giants, gas giants \citep{cleeves2014, drake2005, Hartogh2011}, and brown dwarfs as predicted from \citet{spiegel2011} (dark grey line), with the retrieved D/H measurement of WISE 0855 in black. The mass estimate is derived from the retrieved log(g) with a radius of 1.0 $R_{\rm Jup}$.}
   \label{fig:dtoh}
\end{figure}

The detection of deuterium in WISE 0855, and the inference of a mass below 12 M$_{\rm Jup}$, agrees with theoretical cooling models from \citet{saumonmarley2008} and \citet{phillips2020} which predict a mass between 3 - 10 M$_{\rm Jup}$ for ages between 1 and 10 Gyr. 

\subsection{Prospect for \texorpdfstring{CH$_{3}$D}a detection in exoplanets and brown dwarfs}
\citet{Morley2019} predicted that a R $\sim$ 2700 spectrum with S/N $>$ 40 would be sufficient to detect CH$_{3}$D in a T$_{\rm eff}$ = 300 K atmosphere. Here we have shown that CH$_{3}$D is detectable in R $\sim$ 1000 spectra with S/N $>$ 100. CH$_{3}$D should be detectable in cold, isolated Y dwarfs in this mass range with moderate resolution spectra from JWST at the above S/N.

We removed all CH$_{3}$D opacity from the best fit model to the GO data set to determine its impact on G395M spectra and to determine the spectral S/N needed to detect CH$_{3}$D. The removal of CH$_{3}$D affected the flux most strongly between 4.31 $\mu$m and 4.67 $\mu$m. Its removal caused a flux difference $>$0.5$\%$ (corresponding to an S/N = 200) in 58 pixels, $>$1$\%$ (S/N = 100) in 20 pixels, and $>$1.5$\%$ (S/N=67) in 10 pixels. Other molecular opacity sources in the 4.31 — 4.67 $\mu$m range include CO$_{2}$ at the bluer wavelengths and CO at the redder wavelengths. As CO becomes the dominant carbon reservoir (through a hotter effective temperature or more vigorous vertical mixing), CH$_{3}$D features become less observable. 

The detection of deuterium in an atmosphere outside of the solar system with \textit{JWST} opens several intriguing possibilities for planet characterization and our understanding of planetary formation. The absorption strengths of both CH$_{3}$D and HDO increase relative to those of their non-deuterated counterparts at colder temperatures, and their detection becomes more difficult as the temperature of the planet increases. Cold, faint exoplanets are difficult to observe, but \citet{Molliere2019_isotopes} have shown that CH$_{3}$D is detectable in exoplanets with the Extremely Large Telescope (ELT), and that HDO is potentially detectable in exoplanets with JWST if strong vertical mixing removes most of the CH$_{4}$ from the photosphere. Additionally, \citet{Morley2019} found that CH$_{3}$D is detectable for any cold ($\approx$ 320 K), young ($<$ 20 Myr) Neptunes discovered with JWST. Since deuterium is easier to detect in higher metallicity atmospheres, a S/N ratio of 5 would be sufficient to detect a protosolar abundance of deuterium in a young Neptune-twin ([M/H]=2.0). The D/H ratios as a function of object mass for objects in the solar system, low-mass brown dwarf models, and this work are shown in Figure \ref{fig:dtoh}.

Additionally, a higher D/H ratio, and thus a higher CH$_{3}$D$/$CH$_{4}$ or HDO$/$H$_{2}$O ratio, makes deuterium easier to detect. These ratios would be inflated in an atmosphere that has undergone significant atmospheric mass loss that preferentially loses the lighter hydrogen over the heavier deuterium. \citet{Molliere2019_isotopes} predict that HDO in a non-transiting GJ-1214 b twin at half the distance to earth and D/H ratio of 3 $\times$ 10$^{-4}$ is observable with with the ELT. \citet{cherubim2024} and \citet{Gu2023} have predicted that HDO may be observable for several irradiated, deuterium-enriched sub-Neptunes. 

Unfortunately, a large portion of the most detectable HDO band (3.6 $\mu$m $<$ $\lambda$ $<$ 4.0 $\mu$m) as described in \citet{Molliere2019_isotopes}, \citet{Morley2019}, and \citet{chabrier2000} is in the G395H/S200A1 detector gap (3.69 $\mu$m $<$ $\lambda$ $<$ 3.79 $\mu$m) or the G395H/S200A2 detector gap (3.81 $\mu$m $<$ $\lambda$ $<$ 3.92). However \citet{Molliere2019_isotopes} determined HDO is still detectable with only 3.60 -- 3.80 $\mu$m spectra, so G395H/S200A2 may be preferred if using the highest resolution mode of JWST. G395M, the lower resolution mode of JWST/NIRSpec, does not have this gap and may be more useful in capturing the entire HDO band. 

\subsection{Isotopologues and their potential for tracing planet formation}
Beginning with the detections of $^{13}$CO in a directly imaged, widely separated substellar object TYC 8998 b by \citet{zhangy2021_isotopejupiter} and in an isolated brown dwarf 2MASS J03552337+1133437 by \citet{zhangy2021_isotopebd}, isotopologue detections in brown dwarf and exoplanet atmospheres have become possible due to new ground- and space- based moderate and high resolution infrared spectrographs. Subsequent $^{12}$CO/$^{13}$CO has been determined (or had an upper limit determined) for several brown dwarfs \citep{Hood2024, xuan2024,lew2024} and exoplanets \citep{line2021Natur,Finnerty2023,  gandhi2023, Finnerty2024,smith2024,xuan2024_companions,gonzalez2024}, with ratios appearing lower for objects that formed in disks compared to those that did not. JWST has enabled detections of other isotopologues such as $^{15}$NH$_{3}$ in \citet{barrado2023}, C$^{17}$O and/or C$^{18}$O in \citet{gandhi2023} and \citet{gonzalez2024}. This work is the first to detect the isotope deuterium in an extrasolar atmosphere. 

Isotopologues, particularly those involving $^{13}$C and deuterium, have been detected in disks for decades \citep{Dutrey1994,Guilloteau2013,bergin2013,Huang2017, tobin2023,podio2024}. The differing optical depths of $^{13}$CO and $^{12}$CO allow them to be used to probe different parts of the disk. If future disk observations and modeling can determine isotoplogue ratios as a function of separation, they can act as a planet formation and migration tracer alongside traditional tracers like carbon and oxygen. 

\subsection{\texorpdfstring{PH$_{3}$}a and chemical disequilibrium}
\begin{figure}
    \centering
    \includegraphics[width=1.0\linewidth]{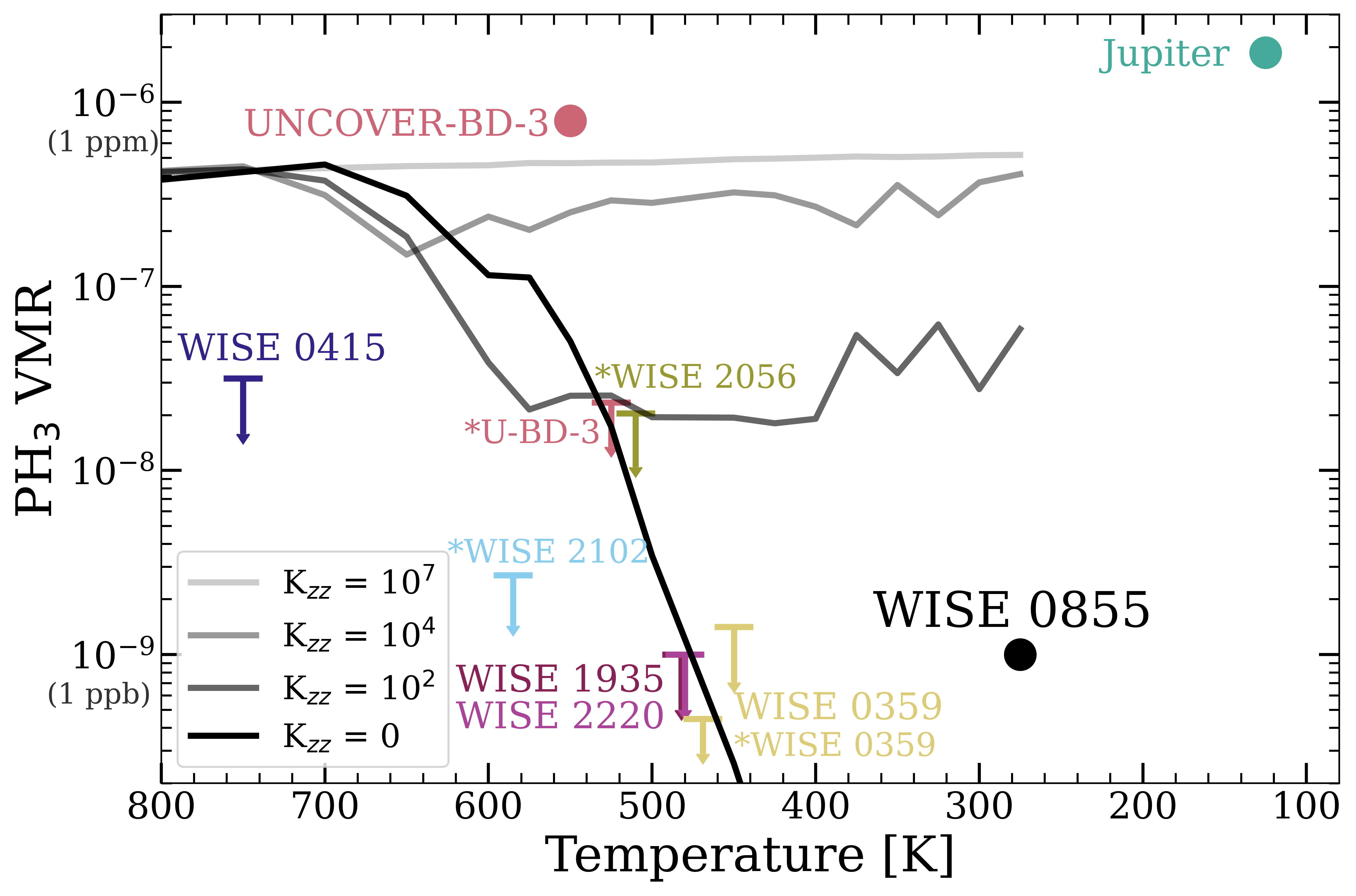}
    \caption{The PH$_{3}$ abundance as a function of T$_{\rm eff}$ for Y dwarfs with detections or upper limits from \citet{Hood2024}, \citet{burgasser2024}, \citet{faherty2024}, \citet{kothari2024}, \citet{Beiler2024a}[*], and this work, along with the global PH$_{3}$ abundance of Jupiter from \citet{fletcher2009}. The grey and black lines show the Elf Owl PH$_{3}$ abundance at 3 bars for various mixing strengths and effective temperatures. All models assume log(g) = 4, cloud-free, solar metallicity, and solar C/O atmospheres. Models extend down to 275 K. WISE 0855 is consistent with very slow (log(K$_{\rm zz}$) $<$ 2 cm$^{2}$s$^{-1}$) mixing.}
   \label{fig:ph3}
\end{figure}

We detect approximately 1 ppb of phosphine (PH$_{3}$) in the atmosphere of WISE 0855. At the cold temperature of this Y dwarf, we expect the presence of disequilibrium gas species with abundances quenched at higher values found deeper in the atmosphere. While the rest of the gas abundances are consistent with rapid vertical mixing of log(K$_{\rm zz}$) $>$ 7 cm$^{2}$s$^{-1}$, the low abundance of PH$_{3}$ appears consistent with chemical equilibrium or very slow mixing (log(K$_{\rm zz}$) $<$ 2). The abundance of PH$_{3}$ reported here, the previously reported non-detections of PH$_{3}$, and predicted abundances from \citet{mukherjee2022} as a function of T$_{\rm eff}$ and log(K$_{\rm zz}$) are shown in Figure \ref{fig:ph3}. 

Our measured PH$_{3}$ abundance of 10$^{-8.91}$ in the previously published \citet{Luhman2024} data set and 10$^{-9.24}$ in the GO data set are consistent with the modeling in \citet{Luhman2024} which found that a PH$_{3}$ abundance of 10$^{-8}$ produced too strong an absorption feature to fit their data. Based on L band (3.4 - 4.14 $\mu$m) and M band (4.5 - 5.1 $\mu$m) spectra from \citet{skemer2016}, \citet{Morley2018} placed an upper limit of PH$_{3}$ of 10$^{-6.30}$ for this object. \citet{miles2020} found an underabundance of PH$_{3}$ in WISE 0855 compared to the amount predicted by the vertical mixing timescale inferred from CO abundances (log(K$_{\rm zz}$) = 8.5).

Our log(K$_{zz}$) values inferred from our retrieved CO abundance ($>$ 7) and PH$_{3}$ abundance ($<$ 2) are discrepant. Our inference of slow vertical mixing depends on the accuracy of the phosphorus thermochemical timescales of \citet{Gurvich1989}. The number of non-detections shown in Figure \ref{fig:ph3} and the large number of non-detections in hotter T dwarfs not shown indicate that our knowledge of phosphorous chemistry is incomplete. The low but constrained abundance in WISE 0855 adds information that will hopefully aid in the identification of problematic elements in the phosphorous chemical reaction timescale network. 

\citet{burgasser2024} made the first claim of PH$_{3}$ in an extra-solar atmosphere based on 4.2 $\mu$m absorption in the NIRSpec/PRISM spectra of UNCOVER-BD-3, a cold (T$_{\rm eff}$ = 550 K), low-metallicity ([M/H] = -1.0) Y-dwarf with potential halo membership. Recent work by \citet{beiler2024b} highlighted the difficulty in disentangling PH$_{3}$ absorption from CO$_{2}$ with forward model fitting due to the overabundance of PH$_{3}$ and underabundance of CO$_{2}$ in current models. They were not able to rule out PH$_{3}$ in UNCOVER-BD-3, but found the 4.2 $\mu$m feature could be explained by excess CO$_{2}$ compared to model predictions. 

\section{Conclusions}\label{Conclusions}
Here we report a detection of a deuterated gas species, CH$_{3}$D, and a part-per-billion level abundance of PH$_{3}$ in the atmosphere of the coldest brown dwarf, WISE 0855. These detections highlight the extreme sensitivity of JWST. Our CH$_{3}$D and PH$_{3}$ abundances were retrieved in two independent data sets reduced by different groups and were robust to a variety of model assumptions. Models with CH$_{3}$D and PH$_{3}$ opacities were strongly preferred over models without them. An injection and retrieval test indicates that the CH$_{3}$D information present in NIRSpec/G395M spectra is sufficient to constrain the CH$_{3}$D abundance. A cross correlation analysis of the residuals detected PH$_{3}$ at a CCF S/N of 8.5. Differences in the posteriors of other parameters may be driven by single-epoch vs averaged time series observations and will be explored in a future work. 

We derive a protosolar D/H ratio from an abundance of CH$_{3}$D retrieved in JWST NIRSpec/G395M spectra. The presence of deuterium in the atmosphere of a free-floating object indicate that WISE 0855 has a mass less than the deuterium-burning limit of $\sim$ 12 M$_{\rm Jup}$, in agreement with evolutionary models. It also underscores the opportunity JWST presents to find this powerful formation tracer in exoplanet atmospheres, as predicted by \citet{Molliere2019_isotopes} and \citet{Morley2019}. 

\section{Acknowledgements}
\textit{Software}: CHIMERA \citep{line2013b}, \texttt{emcee} \citep{foremanmackey2013}, \texttt{corner.py} \citep{foremanmackey2016}

We would like to acknowledge Elena Manjavacas for her expertise and advice on the ESA and STSci pipelines. We would like to acknowledge Stephan Birkmann for advice on the ESA pipeline and providing intermediate data products. We would like to acknowledge the JWST Helpdesk Team, 
 and Lonestar6 at the Texas Advanced Computing Center (TACC) at The University of Texas at Austin. 
 Support for programs \# JWST-AR-01977.004 and JWST-GO-02124.009-A were provided by NASA through a grant from the Space Telescope Science Institute, which is operated by the Associations of Universities for Research in Astronomy, Incorporated, under NASA contract NAS5- 26555.
 This material is based on work supported by the National Aeronautics and Space Administration under grant No. 80NSSC21K0650 for the NNH20ZDA001N-ADAP:D.2 program. 
 M.J.R. acknowledges funding from NASA FINESST grant No. 80NSSC20K1550.
 C.V.M. acknowledges support from the Alfred P. Sloan Foundation under grant number FG-2021-16592.
 C.V.M. acknowledges the National Science Foundation, which supported the work presented here under Grant No. 1910969.
 B.E.M. was supported by the Heising–Simons Foundation 51 Pegasi b Postdoctoral Fellowship. 
 J.F. acknowledges funding from the Heising Simons Foundation as well as NSF award \#2238468, \#1909776, and NASA Award \#80NSSC22K0142
 J. M. V. acknowledges support from a Royal Society - Science Foundation Ireland University Research Fellowship (URF$\backslash$R1$\backslash$221932). 
B.B. acknowledges support from UK Research and Innovation Science and Technology Facilities Council [ST/X001091/1].
R.L acknowledges support from STScI Grant number JWST-GO-02327.

\appendix
\section{Data Set Comparison}\label{AppendixA}

The \cite{Luhman2024} spectrum was reduced using a custom pipeline developed by the European Space Agency NIRSpec science operations team, which adopts the same algorithms included in the standard JWST pipeline, except for a correction for ``snowballs” and a correction for residual correlated noise\footnote{\url{https://jwst-tools.cosmos.esa.int/}}{\citep{2018SPIE10704E..0QA, 2022A&A...661A..81F, 2023PASP..135c8001B}. The median percent difference of the \cite{Luhman2024} spectrum to our spectrum between 4.00-5.09 $\mu$m is on average 4$\%$, with a standard deviation of 13$\%$. We show the differences between both reductions over the region where CH$_{3}$D is detected in Figure~\ref{fig:spectra_comparison}. Most of the differences seen between the two reductions can be explained by the inherent variability in the GO data. We believe the outliers to be a result of different data reduction pipelines, which is beyond the scope of this paper. The data from the GTO program were also re-reduced using the standard JWST Pipeline through stages 1-3 with the same CRDS version and context as the time-series data. When using the same raw detector files, the median ratio between the spectrum produced by the NIRSpec science operations team pipeline and the standard JWST pipeline is 11$\%$ with a standard deviation of 14$\%$. 

\begin{figure}
    \centering
    \includegraphics[width=1\linewidth]{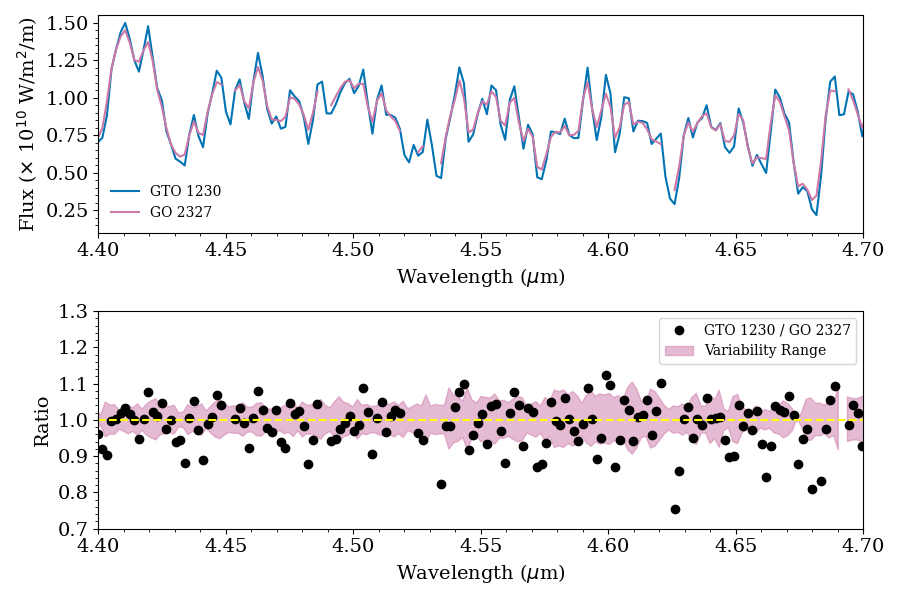}
    \caption{Comparison of reduced spectra within the region most impacted by CH$_{3}$D absorption. Top panel: The GTO spectrum [blue] plotted with the GO spectrum [pink]. The GO spectrum is interpolated onto the same wavelength grid as the GTO spectrum. Bottom panel: The ratio of the GTO spectrum divided by the GO spectrum [black dots]. The yellow dashed line is a reference for no difference. The top of the pink area is the ratio of the brightest spectrum divided by the dimmest spectrum in the time-series. The bottom of the pink area is the dimmest spectrum divided by the brightest spectrum.} 
   \label{fig:spectra_comparison}
\end{figure}




\section{List of Retrievals and Retrieved Parameters}\label{AppendixB}
In most retrievals, model spectra were rotationally broadened using the \texttt{fastRotBroad} function from \texttt{PyAstronomy} as in \citet{Hood2023} and \citet{Hood2024} \citep{Czesla2019}. \texttt{fastRotBroad} was found by \citet{xuan2024} to be invalid over a large wavelength range, so additional retrievals were performed using \texttt{RotBroadInt} from \citet{carvalho2023RNAAS}. Retrievals performed with \texttt{RotBroadInt} took an average of 69.5 h compared to 34.5 h for retrievals with \texttt{fastRotBroad}. The G395M spectra were insufficient to accurately constrain \textit{v} sin \textit{i} values, and the use of the \texttt{RotBroadInt} function did not change other retrieved values. We attribute the discrepancy in the retrieved radial velocities between the two data sets to different wavelength calibrations between the two pipelines.

A subset of retrieved posteriors is provided in Figure \ref{fig:appendix_corner}. Retrieved $\log_{10}$($\frac{\rm CH_{3}D}{\rm CH_{4}}$)$_{E}$ values remained consistent across all retrievals performed on the same data set. The retrieved errorbar inflation parameter was also consistent for all retrievals performed on the same data set. Other bulk properties remained consistent across retrievals with the exception of the non-uniform H$_{2}$O retrieval, which retrieved $\log_{10}$(H$_{2}$O) abundances of -3.12 $\pm$ 0.07 in the deep atmosphere and -3.40 $\pm$ 0.05 in the upper atmosphere. We used the deep atmosphere abundance to calculate [M/H] = -0.06 $\pm$ 0.07 and C/O = 0.24 $\pm$ 0.02. The two additional parameters needed for the nonuniform H$_{2}$O profile did not appreciably improve the fit and the model was not preferred. All cloud models assumed H$_{2}$O Mie scattering clouds. We retrieved an upper limit for the cloud volume mixing ratio (see Figure \ref{fig:appendix_corner}), and the other cloud parameters were unconstrained across the performed retrievals. We do not draw conclusions about the presence or absence of water clouds, but these results indicate that clouds modeled with Mie scattering and log-normal particle size distributions may not be sufficient to fit the spectra of cold objects. 

We followed the method described in \citet{zhangy2021_isotopebd} to perform a cross-correlation analysis to test the detection of CH$_{3}$D and PH$_{3}$. Our cross-correlation function (CCF) used the residuals (the observed spectrum minus the best fit model with CH$_{3}$D or PH$_{3}$ opacity removed) and the CH$_{3}$D or PH$_{3}$ model spectrum. This model spectrum was created by subtracting the best fit spectrum from the same model but with the targeted opacity set to 0. The CCF was then normalized by its standard deviation outside of the peak within the velocity of [-10000, -800] and [800, 10000]. This CCF is shown in Figure \ref{fig:appendix_corner} and results in a CCF S/N of 3.8 for CH$_{3}$D and a CCF S/N of 8.5 for PH$_{3}$. We note the prominent troughs and inconclusive result of the CH$_{3}$D CCF. We attribute this to the narrow wavelength range of the most prominent CH$_{3}$D feature and the relatively low (R$\sim$ 1000) resolution of NIRSpec/G395M. 

\begin{deluxetable}{llc}
\tablecolumns{3}
\tablecaption{Retrieved Parameters\label{tab:params}}
\tablehead{
\colhead{Parameter} & \colhead{Description}& \colhead{Prior}}
\startdata
log$_{10}$(g) & log$_{10}$ of surface gravity [cm s$^{-2}$] & M$<$100 M$_{\rm Jup}$\\
(R/D)$^{2}$ & radius-to-distance scale factor [$R_{\rm Jup}$ pc$^{-1}$] & 0$<$(R/D)$^{2}$$<$1\\
10$^{b}$ & errorbar inflation & 0.01$\times$($\sigma_{min}^{2}$), 100$\times$($\sigma_{min}^{2}$)\\
T$_{i}$ & Temperature [K] at a given pressure level & $<$4000 K \\
$\gamma$ & TP profile smoothing hyperparameter & 0 - $\infty$\\
log$_{10}(f_{i})$ & log$_{10}$ of VMR of a uniform gas & $>$-12, $\sum$f$_{i}$$<$1\\
$\log_{10}$($\frac{\rm CH_{3}D}{\rm CH_{4}}$)$_{E}$ & log$_{10}$ ratio relative to terrestrial (1/1606) & -3 - 3 \\
log$_{10}$(C) &  log$_{10}$ of the cloud volume mixing ratio & -15 - 0\\
log$_{10}$(P$_{c}$) &  log$_{10}$ of the cloud base pressure & -15 - 0\\
f$_{sed}$ & sedimentation efficiency & 0 - 10 \\
RV & radial velocity [km s$^{-1}$] & -50 - 50 \\
\textit{v} sin \textit{i} & rotational velocity [km s$^{-1}$] & 0 - 140\\
\enddata
\end{deluxetable}

\begin{deluxetable}{cccccccccc}
\tablecolumns{10}
\tablecaption{List of Retrievals\label{tab:retrievals}}
\tablehead{
\colhead{Data} & \colhead{Clouds} & \colhead{P(T)} & \colhead{H$_{2}$O} & \colhead{Broadening} & \colhead{CH$_{3}$D} & \colhead{PH$_{3}$} &  \colhead{Parameters} & \colhead{$\log_{10}$($\frac{\rm CH_{3}D}{\rm CH_{4}}$)$_{E}$} & \colhead{$\Delta$ BIC} }
\startdata
GO & yes & smoothed & uniform & FastRotBroad & yes & yes & 36 &-1.09$^{+0.06}_{-0.07}$ & N/A\\
GO & yes & smoothed & uniform & FastRotBroad & no & yes & 35 & N/A & -44.9\\
GO & yes & smoothed & uniform & FastRotBroad & yes & no & 35 &-1.10$^{+0.09}_{-0.10}$  & -27.9\\
\hline
GO & none & smoothed & uniform & FastRotBroad & yes & yes & 33 &-1.10$^{+0.06}_{-0.07}$  & N/A\\
GO & none & smoothed & uniform & FastRotBroad & no & yes & 32 & N/A & -44.1\\
\hline
GO & yes & unsmoothed & uniform & FastRotBroad & yes & yes & 35 &-1.02$^{+0.06}_{-0.06}$  & N/A\\
GO & yes & unsmoothed & uniform & FastRotBroad & no & yes & 34 & N/A & -55.7\\
\hline
GO & yes & smoothed & uniform & RotBroadInt & yes & yes & 36 &-1.09$^{+0.06}_{-0.07}$  & N/A\\
GO & yes & smoothed & uniform & RotBroadInt & no & yes & 35 & N/A & -42.8\\
\hline
GO & yes & smoothed & non-uniform & RotBroadInt & yes & yes & 38 &-1.14$^{+0.10}_{-0.07}$ & N/A\\
GO & yes & smoothed & non-uniform & RotBroadInt & no & yes & 37 & N/A & -18.3\\
\hline
\hline
GTO & yes & smoothed & uniform & FastRotBroad & yes & yes & 36 & -0.94$^{+0.06}_{-0.07}$ & N/A\\
GTO & yes & smoothed & uniform & FastRotBroad & no & yes & 35 & N/A & -55.4\\
GTO & yes & smoothed & uniform & FastRotBroad & yes & no & 35 &-0.92$^{+0.06}_{-0.07}$ & -61.9\\
\enddata
\end{deluxetable}

\begin{figure}
    \centering
    \includegraphics[width=1\linewidth]{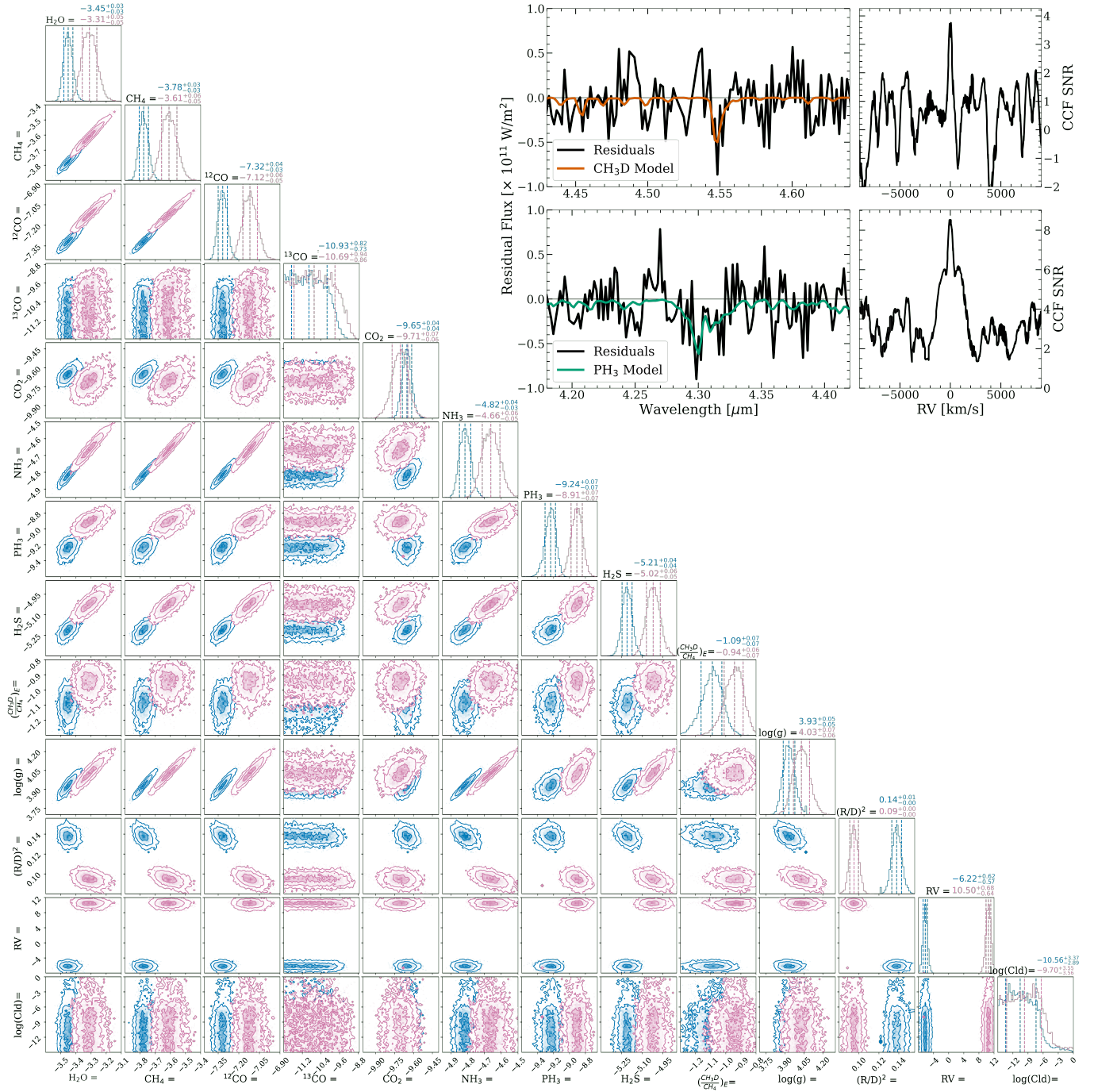}
    \caption{The corner plot of a subset of retrieved parameters from the GTO spectrum [pink] and the time-averaged GO spectrum [blue]. The discrepancy between the retrieved radial velocity are believed to be due to differences in wavelength calibration in the two pipelines. An upper limit for the cloud volume mixing ratio is shown, and other cloud parameters were unconstrained. Top panel: The cross correlation function (CCF) computed for the GO dataset with the CH$_{3}$D model [top] or PH$_{3}$ model [bottom] and residuals (observed spectrum minus the best fit model with the relevant opacity removed). The CCF is normalized by the standard deviation outside of the CCF peak so that the y-axis of the right-hand panels represent the S/N of the CCF peak.}
   \label{fig:appendix_corner}
\end{figure}

\clearpage
\bibliographystyle{aasjournal}
\bibliography{references}

\end{document}